\documentclass[hoptionsi,screen,format=acmsmall]{acmart}
\usepackage{graphicx}				

\usepackage{amsmath}

\usepackage{xypic}
\usepackage{cancel}
\usepackage{mathtools}
\usepackage{tikz-cd}
\usepackage{musicography}

\theoremstyle{definition}
\newtheorem{definition}{Definition}[section]

\newcommand{\Pc}{\mathcal{P}}

\newcommand{\PRb}{\mathit{PR}}
\newcommand{\Relb}{\mathit{Rel}}
\newcommand{\Setb}{\mathit{Set}}

\newcommand{\band}{\mathop{\&}}
\newcommand{\bor}{\mathop{\|}}

\title{The Semantics of Package Management via Event Structures}
\author{Gershom Bazerman}
 \affiliation{%
   \institution{Awake Security}
   \country{United States}}
 \email{gershomb@gmail.com}
 
 \acmBooktitle{N/A} 

\begin{abstract}
We propose an approach to the semantics of package management which relates it to general event structures, well-known mathematical objects used in the semantics of concurrent, nondeterministic systems. In this approach, the data of a package repository is treated as a declarative specification of a nondeterministic, concurrent program. We introduce a process calculus corresponding to this data, and investigate its operational and categorical semantics. Our hope is this lays the basis for further formal study of package management in which the weight of existing tools can be brought to bear.

\end{abstract}

\begin{document}
\maketitle

\section{Introduction}
Package management is an understudied problem in computer science. Yeoman's work has been done in \cite{di2006edos, abate2012dependency}, where the problem of solving dependencies of a package was shown to be NP-complete. But generally, the task of package management has been seen as one of engineering, and not of formal specification and analysis. 

As described in \cite{abate2012dependency}, addressing package management for linux distributions such as Debian:
\begin{quotation}
Two related phenomena, the explosion of Internet connectivity and the mainstream adoption of free and open source software (FOSS), have deeply changed the scenarii that today’s software engineers face. The traditional organized and safe world where software is developed from specifications in a fully centralized way is no longer the only game in town. We see more and more complex software systems that are assembled from loosely coupled sources developed by programming teams not belonging to any single company, cooperating only through fast Internet connections. The availability of code distributed under FOSS licences makes it possible to reuse such code without formal agreements among companies, and without any form of central authority that coordinates this burgeoning activity.
\end{quotation}

The same holds true for package management for particular software build and distribution chains within any given language. As such, package management, once an afterthought, is now a central element in the development of modern large-scale software. We believe that package management for programming languages can and should be approached with the same care and precision that the specification of the "internal" components of languages themselves is given.

Our aim is to shed some light on the essence of package management, showing that it takes semantics in structures familiar from the study of concurrent programs, and hence opening it up to future study from a programming languages perspective. The general intuition, which this paper will make rigorous, is that the metadata of a given package repository corresponds to the declarative specification of a concurrent, nondeterministic program. Each package may be treated as an ``event'' in the program, which declares the enabling conditions for it to occur. An end-user who installs packages from the repository plays the role of the scheduler, determining which choices are to be made, and in what order. Further, a maintainer adding a new package to the repository is specifying a conservative program transformation. Thus, the history of an end-user's interaction with the repository -- installing packages from it -- is considered as a trace of a single execution of the program, and a maintainer's interaction with the repository (adding new packages) corresponds to monotone extension of the program into a new program.

The process calculus we introduce for this also turns out to serve as a convenient algebraic notation for the data of general event structures, as used in concurrency theory. We also introduce a category of such structures, and show how it may be used to give semantics to this calculus, and, transitively, to basic operations on package repositories.

As an application, we show how these semantics may let us distinguish between various models of package management, and also how they let us formalize certain notions of a ``package version policy''.

\section{Packages Management and Event Structures}
First we introduce some terminology which abstracts across many package repositories (npm, hackage, RubyGems.org, Maven Central, etc.) and the details of their package management tools. A \textbf{package} consists of either raw source code or compiled artifacts, coupled with \textbf{metadata}, which is information that is not directly used by the compiler or interpreter, but instead describes information used in configuring or planning a build, or a sequence of builds. Metadata is typically   given in a file such as \texttt{package.json},  \texttt{pom.xml}, or \texttt{packagename.cabal}. The most salient metadata is the distinguishing identification of the package (typically name and version) as well as the dependencies necessary for the package to function or build properly (often with the possibility that there may not be a fixed set of dependencies, but rather a logical expression indicating various alternative choices of such).

As discussed, a given package name is typically coupled with a \textbf{package version}, which, for simplicity's sake, we will always consider as a pair of integers, separated by a dot, so \texttt{foo-3.2} is the package ``foo'' with the major version 3, and minor version 2. (In practice, many languages have a third component and sometimes fourth component which is often used to specify a ``patch'' level bugfix-only release, but treating that in this paper would only clutter the exposition while yielding no new insight). For the initial part of this paper, we will consider two packages with the same name but different versions as entirely different packages. Eventually, we will describe an approach to recover the association between different versions of the same package name.

A \textbf{package repository} is a collection of packages, which maps package identifiers to the underlying content they index, and typically with a mechanism to fetch the entirety of the metadata from all packages. Since this paper does not yet address multiple interaction package repositories, we make the simplifying assumption that package repositories are closed under dependency. This is to say any package in a repository will necessarily specify as dependencies either ``base'' libraries bundled with a given programming language, or other packages also in the repository. 

Package repositories come with some associated operations. One can always \textbf{initialize} a package repository in an empty state with no packages. One can also \textbf{extend} a package repository through the addition of a new package, typically through a command such as \texttt{gem push}, \texttt{npm publish} or \texttt{cabal upload}. \cite{npm, cabal, gem} As a simplifying assumption in this paper, we treat the case when the extension of a repository through a new package only occurs when all dependencies for that package already exist in the repository.

Finally, there are \textbf{package managers}, which are command-line tools such as \texttt{npm}, \texttt{cabal}, \texttt{gem}. We define an \textbf{environment} to be a collection of installed packages, and treat installation as simply the addition of a new package to this environment. In the basic model we present, installation to an environment can only occur when necessary dependencies for a package already exist in the environment. A package manager is then just a tool which manages installation of packages into an environment.

\subsection{General Event Structures}

We recall the algebraic definition of a general event structure, as given in \cite{winskel1986event}, and describe how it is to be interpreted.

\begin{definition}
A \textbf{general event structure} is a triple \((E, Con \subset Fin(E), \vdash \, \subset Con \times E)\) such that

 \renewcommand{\labelenumi}{\roman{enumi}}
  \begin{enumerate}
   \item \(E\) is a set of events.
   \item \(Con\) is the consistency predicate (or collection of contexts) given as a downward-closed collection of finite subsets of \(E\). This is to say that if any set is an element of \(Con\), then all subsets of that set are also elements of \(Con\).
   \item \(\vdash\) is the enabling relation, such that if some context enables some event, then all larger contexts also enable that event.
 \end{enumerate}
\end{definition}

It is straightforward to interpret this as the specification of a concurrent process. A process specification consists of some events that can happen, given by the set \(E\), and some preconditions on events, which indicates that some events may only occur after others have enabled them, given by the enabling relation \(\vdash\). The condition on the enabling relation ensures that it behaves in this manner, specifying that any event which is enabled by some set of events can likewise be ensured by any set of events containing that first set. While enabling relation is defined in this "upwards-closed" fashion, it could equally well be given simply as generated by a relation between events and minimal enabling contexts -- i.e. sets which enable the event, and for which any subset does not. It is also worth nothing that if we had not a relation, but a simple mapping \(E \to Con\), then every event would have a single unique minimal enabling context. Such a structure is known simply as an ``event structure'' -- i.e., the ``general'' in ``general event structure'' refers precisely to the fact that events may be enabled in multiple fashions.

Finally, just as certain events may require others to occur first, some events may prevent others from occurring, i.e. they may conflict. For example, two events may consume the same resource, in which case if one occurs, then the other cannot. This is the information captured by the consistency predicate, \(Con\), which picks out only consistent events. When some collection of events cannot occur together in any context, we say that it is in \textbf{conflict}, or that the events within it are \textbf{inconsistent}. The downward-closure condition ensures that any sub-context of a context with no conflicts also has no conflicts. While the consistency predicate is given as a collection of consistent contexts, it could equally well be given by a generating collection of minimal inconsistent contexts -- i.e., pairs, triples, etc. of events such that they are inconsistent but any subset is consistent.

\subsection{The Data of Package Repositories}

We now consider in more detail the metadata of package repositories. Below is an example \texttt{package.json} file used to describe JavaScript packages in the npm package repository:

\begin{verbatim}
{
  "name": "leftpad", 
  "version": "5.9.2", 
  "description": "Provides left padding",
  "main": "index.js",
  "license": "MIT",
  "dependencies": {
      "react": "2.4.0",
      "webpack": "0.1.3",
      "redis": "^4.3.0"
   }
}
\end{verbatim}

The name and version properties serve to uniquely identify the package within a given repository. The description and license are intended, for the most part, for human consumption, and we need not consider them here. The ``main'' property is used to indicate the entry point to the package to the build system, but we are not concerned with the actual build process at this moment, so it need not be considered either. The central thing to understand is then the ``dependencies'' property. Here, it is given as a mapping of package names to package version specifications. So the \texttt{react} package is required to already be available at precisely version 2.4.0, and similarly for \texttt{webpack}. On the other hand, a caret precedes the version specification for \texttt{redis}. In the syntax of these files, this indicates that any version of the package with major version 4 and minor version >= 3 is acceptable. I.e. it specifies not just a single version, but a range. Supposing we knew for a fact that the matching versions were 4.3.0, 4.4.0 and 4.5.0, then the meaning of this dependency would be the disjunctive clause: \texttt{redis-4.3.0 $\bor$ redis-4.4.0 $\bor$ redis-4.5.0}. Further, since each package gives a disjunction of versions, but the dependency field as a whole specifies a conjunction of packages, then the entirety of the key metadata for a package may be regarded as a single implicative formula:

\smallskip
\texttt{
leftpad-5.9.2 $\to$ react-2.4.0 $\band$ webpack-0.1.3 $\band$ (redis-4.3.0 $\bor$ redis-4.4.0 $\bor$ redis-4.5.0)
}
\smallskip

We may perform the same analysis on metadata files for other languages and package management systems. A \texttt{cabal} file that specifies a Haskell package has, per-component (since packages may provide libraries, multiple executables, tests, etc.), a ``build-depends'' field, like the following:

\begin{verbatim}
  build-depends:
    base             >= 4.7.0.0 && < 5,
    bytestring       >= 0.10.4.0 && < 0.11,
    containers       >= 0.5.5.1 && < 0.7
\end{verbatim}

Again, within a fixed universe of packages, we can now ``extract'' the core logical formula of package metadata as an implication between a packagename-component-version triple and an expression in conjunctive normal form. It is worth noting that arbitrary disjunctions and conjunctions may be specified in cabal version ranges. Further, more complex cabal files may contain flags so that richer expressions are possible -- i.e. a package could depend on one of two providers of matrix computations, or be built with either a gui or console frontend, or the like. In all such cases, the dependencies can of course still be boiled down to an expression solely involving conjunction and disjunction.

We will not repeat the same exercise for Ruby gemfiles, or Maven \texttt{pom.xml} files or the like, but things work out very similarly.

\subsection{The Correspondence}

As we have seen, for our purposes, the metadata of a repository can be given entirely as a collection of these implications, each mapping a single package identifier to an expression in positive (i.e. negation-free) propositional logic. We will show that this gives exactly the data of a (conflict-free) general event structure.

At a high level, all we need to is read ``event set'' as ``package set'', take the consistency predicate to be the full powerset of packages, and read ``enabling relation'' as ``dependency relation''. 

The key insight, insofar as any is necessary, is that the enabling relation, while given as a relation between events (resp. packages) and enabling contexts (resp. dependency sets) can equally be given as a mapping function between events and sets of sets of events. I.e. \(a \, R \, b \simeq a \to \Pc(b)\). Hence an enabling relation \(E, R, Fin(E)\) can be alternatively written as a mapping \(E \to \Pc(Fin(E))\).

In turn, the double-powerset of a set can be interpreted as a logical formula over that set in disjunctive normal form -- i.e. as a disjunction of inner conjunctive clauses over the atoms of the set. E.g. \(\{\{a,b\},\{c,d\}\}\) interprets as \(a \band b \bor c \band d\). Since package dependency specifications can all be interpreted in terms of disjunction and conjunction, they can all be appropriately normalized to match the appropriate type.

As described, we have not taken into account conflict at all -- and indeed, it may seem that conflict is not represented in metadata. This is unfortunate, as two packages may actually conflict -- for example, by providing different modules which share the same name. In such a case, when a user tries to bring the module into scope, the compiler may not know what to do (at least in some languages -- all of this of course varies wildly between language ecosystems). In many languages, this is reflected in a requirement -- enforced either in the compiler or in the build system -- that two packages with the same name may not be in use at once. In such cases, we do have a source of data for the conflict relation -- each version of a package is in conflict with all other versions of a package, and thus the consistency predicate is generated as the greatest subset of \(Fin(E)\) such that no context contains the same package at two different versions. \footnote{This suggests that it might be useful to enrich package repositories with other forms of conflict data as well, when such is available or could be calculated}.

To make this entirely clear, we give a definition of package repository metadata such that the correspondence with general event structures can be "read off".

\begin{definition}
Package repository metadata is a triple \((P, Con \subset Fin(P), \mapsto : E \to \Pc(Fin(E)))\) such that

 \renewcommand{\labelenumi}{\roman{enumi}}
  \begin{enumerate}
   \item \(P\) is a set of packages.
   \item \(Con\) is the consistency predicate (or collection of contexts) given as a downward-closed collection of finite subsets of \(P\), consisting of all sets of packages that are not in conflict.
   \item \(\mapsto\) is the dependency mapping, which sends packages to minimal possible dependency sets (i.e. sets of enabling dependencies such that none contains another such set). 
 \end{enumerate}
\end{definition}

From the discussion above (and the observation that enabling relations may be represented by their minimal elements), it is clear that this is the same structure.

Since the data of a package repository, as we have now seen, corresponds to the specification of a concurrent program, it is worth considering the intuition that should arise. A single ``execution'' of the program given by a package repository corresponds to starting with the empty environment, and a user installing packages into this environment in some allowable sequence, i.e. ``firing enabled events''. Concurrency corresponds to the fact that a user might set off building and installing two packages at once, assuming they do not depend on one another (or even that a package manager might itself co-ordinate such concurrent builds). Nondeterminism corresponds to the fact that not every package that \textit{can} be installed \textit{is} installed -- i.e. branching choices occur as to which enabled events actually occur.

\subsection{A Concrete Example}

We conclude this section with a concrete example of a small package repository and its interpretation as a general event structure. Below is a cut-down version of four cabal files, corresponding to two packages, each with two versions, such that the latter package depends on either version of the former:

\begin{verbatim}
Name: text
Version: 1
  build-depends: 
---
Name: text
Version: 2
  build-depends: 
---
Name: leftpad
Version: 1.1
  build-depends: 
    text = 1 || 2
---
Name: leftpad
Version: 1.2
  build-depends: 
    text = 1 || 2
\end{verbatim}

Here is the event structure that corresponds to it :
\begin{equation*}
\begin{aligned}
E  = & \{\mathrm{text\text{-}1},\mathrm{text\text{-}2},\mathrm{leftpad\text{-}1.1},\mathrm{leftpad\text{-}1.2}\}
\\\\
Con = & \Pc(\{\mathrm{text\text{-}1},\mathrm{leftpad\text{-}1.1}\}) \cup \Pc(\{\mathrm{text\text{-}1},\mathrm{leftpad\text{-}1.2}\}) \cup
\\ &\Pc(\{\mathrm{text\text{-}2},\mathrm{leftpad\text{-}1.1}\}) \cup \Pc(\{\mathrm{text\text{-}2},\mathrm{leftpad\text{-}1.2}\})
\\\\
\vdash such \, that. 
\\ & \{\mathrm{text\text{-}1}\} \vdash \mathrm{leftpad\text{-}1.1}, \{\mathrm{text\text{-}1}\} \vdash \mathrm{leftpad\text{-}1.2}, 
\\ & \{\mathrm{text\text{-}2}\} \vdash \mathrm{leftpad\text{-}1.1}, \{\mathrm{text\text{-}2}\} \vdash \mathrm{leftpad\text{-}1.2}
\end{aligned}
\end{equation*}

Note that in the above, \(\vdash\) is specified by a \textit{minimal} set of enabling relations. In particular, by the definition of event structures, whenever some valid context enables some event, then every greater valid context also must do so.

\section{Package Repositories as a Programming Language}


Since general event structures give semantics to concurrent programming languages, we should be able to construct a process calculus which corresponds closely with them. The language \textbf{CEP} (Concurrent Exclusionary Processes) is given by the following grammar, where \(e, e_1, e_2\) are drawn from set of variables representing events:

\begin{equation*}
P,Q ::= e![e_1,e_2..].P \bor \overline{e}.P  \bor P + Q \bor 0
\end{equation*}

A term \(e![e_1,e_2].P\) represents an globally exclusive firing an event \(e\) with regards to a (possibly empty) exclusion list. If an event in the exclusion list has occurred, the process blocks. Otherwise, \(e\) fires (if it has not already fired) and furthermore all events in the exclusion list are blocked from being fired. Finally, the program proceeds with process \(P\). When the exclusion list is empty, we abbreviate \(e![].P\) as \(e.P\).

A term \(\overline{e}.P\) represents waiting for \(e\) to fire before proceeding with P. A term \(P + Q\) represents parallel execution of \(P\) and \(Q\). For notational convenience, we write a single bar over a sequence of multiple waits. Finally, \(0\) is the ``nullary'' process which immediately halts. When it is clear for context, we omit a \(.0\) from the end of a firing sequence, so \(a.b.c.0\) is abbreviated as \(a.b.c\). Purely as a notational convenience, we also may consider 0 as an event, so that \(0.a\) represents a process that stops before it can fire \(a\).

This calculus is not as expressive as, e.g., the \(\pi\)-calculus \cite{milner1992calculus}, as there is no theory of local names or replication. In general, there are a few other differences from conventional process calculi. Most notably, there is a notion of global exclusion, and also there is no immediate notion of nondeterministic choice. However, global exclusion is expressive enough to recover nondeterministic choice. In particular, the nondeterministic choice to fire either \(e_1\) or \(e_2\) can be rendered using parallel composition plus exclusion as \(e_1 + e_2![e_1]\). Note that we do not need to add an explicit exclusion to \(e_1\), as if \(e_2\) fires first, then this will already prevent \(e_1\) from firing.

\subsection{Reduction Semantics}

We give reduction semantics making use of a split global context (initially empty) consisting of an ordered list of events which have fired (\(\Gamma\), and of events which are excluded (\(\Delta\)). (In such lists, we append not to the beginning, but to the end.)

\begin{align*}
& \Gamma . e_1 \notin \Gamma \band e_2 \notin \Gamma .. \, , e \notin \Delta \vdash e![e_1,e_2..].P
\\ & \Longrightarrow \\
& \Gamma : e, \Delta : e_1:e_2.. \vdash P
\tag{Firing}
\\\\
& e \in \Gamma , \Delta \vdash \overline{e}.P
\\ & \Longrightarrow \\
& e \in \Gamma , \Delta \vdash P
\tag{Waiting}
\\\\
& \Gamma , \Delta \vdash P + Q
\\ & \Longrightarrow \\
& \Gamma ' , \Delta ' \vdash P' + Q
\\ & where
\\ & \Gamma, \Delta  \vdash P \Longrightarrow \Gamma ' , \Delta ' \vdash P'
\tag{Composition-1}
\\\\
& \Gamma , \Delta \vdash P + Q
\\ & \Longrightarrow \\
& \Gamma ' , \Delta ' \vdash P + Q'
\\ & where
\\ & \Gamma, \Delta  \vdash Q \Longrightarrow \Gamma ' , \Delta ' \vdash Q'
\tag{Composition-2}
\\\\
& \Gamma , \Delta \vdash P + 0
\\ & \Longrightarrow \\
& \Gamma , \Delta \vdash P
\tag{Unit}
\end{align*}

The composition rules give nondeterministic parallel execution. This is to say that the there is choice as to which possible event to fire next, and this can lead to nonconfluent reductions. It would be possible to add a rule that sent processes prefixed with excluded events to \(0\), but even with such a rule it is not the case that all reductions would terminate in a single \(0\) term. For example, consider the deadlock equation \(\overline{a}.b + \overline{b}.a\). 

A relatively standard trace semantics \cite{bloom1995bisimulation, winskel1984synchronization} of a process calculus are then given by taking the synchronization trees induced by the prefix-closed set of every possible \(\Gamma\) that can exist at some stage of some reduction sequence. E.g. the process \(a.b + \overline{a}.c\) has as possible contexts the prefix-closure of \(\{[a,b,c],[a,c,b]\}\) as pictured below:

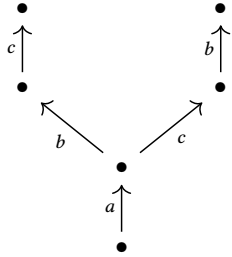
\begin{equation*}
\begin{tikzcd}
\bullet                &                                                & \bullet                \\
\bullet \arrow[u, "c"] &                                                & \bullet \arrow[u, "b"] \\
                             & \bullet \arrow[lu, "b"] \arrow[ru, "c"'] &                              \\
                             & \bullet \arrow[u, "a"]                   &                             
\end{tikzcd}
\end{equation*}

When events are read as ``package installations'' this language can take operational semantics as describing the behavior of a package manager using a given package repository to install a sequence of packages into an environment. The interpretation is, loosely speaking, ``in parallel, try to install everything, blocking when dependencies are not installed, and failing when an incompatible package is first-past-the-post''. In this semantics, the state of package installations in an environment will necessarily correspond to a partial trace of events in some ``execution'' of the program given by the repository, and the collection of installed packages will correspond to the context of fired events.

Our example from the previous section of two versions of \texttt{leftpad} depending on two versions of \texttt{text} now can be translated into CEP as follows:

\begin{align*}
& \mathrm{text\text{-}1} + \mathrm{text\text{-}2}![\mathrm{text\text{-}1}] + 
\\ 
& \overline{\mathrm{text\text{-}1}}.\mathrm{leftpad\text{-}1.1} + 
\\ 
& \overline{\mathrm{text\text{-}1}}.\mathrm{leftpad\text{-}1.2}![\mathrm{leftpad\text{-}1.1}] + 
 \\
& \overline{\mathrm{text\text{-}2}}.\mathrm{leftpad\text{-}1.1} + 
\\ 
& \overline{\mathrm{text\text{-}2}}.\mathrm{leftpad\text{-}1.2}![\mathrm{leftpad\text{-}1.1}] 
\end{align*}

The corresponding synchronization tree (with labels abbreviated) is then:

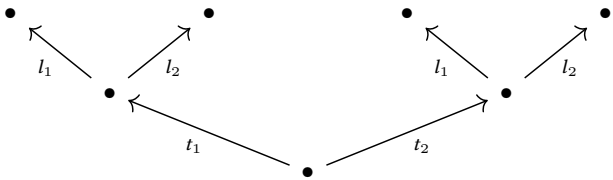
\begin{equation*}
\begin{tikzcd}
\bullet &                                              & \bullet &                                                & \bullet &                                              & \bullet \\
        & \bullet \arrow[lu, "l_1"] \arrow[ru, "l_2"'] &         &                                                &         & \bullet \arrow[lu, "l_1"] \arrow[ru, "l_2"'] &         \\
        &                                              &         & \bullet \arrow[llu, "t_1"] \arrow[rru, "t_2"'] &         &                                              &        
\end{tikzcd}
\end{equation*}

\subsection{Algebraic Properties of CEP}

Synchronization trees give us a notion (or in fact range of notions) of equivalence between processes in CEP. We consider in this section operations that are invariant with regards to the synchronization tree produced, which yield a rich algebraic structure.



The first observation is that parallel composition is associative and commutative, as the two rules for reduction of composition are symmetric. I. Further, it is the case that \(P + P = P\), as any events fired in the first component mean that the related events in the second component are no-ops, and furthermore any wait in the first component is satisfied when the related wait in the second component is (and vice versa). Thus, the synchronization tree produced by \(P + P\) is the same as the tree produced by \(P\) alone. Hence, parallel composition gives an idempotent monoid with \(0\) as the unit

Composition also interacts well with other operations. Firing (including exclusive) and waiting distribute over parallel composition. This is to say that  \(e.(P + Q) = e.P + e.Q\) and \(\overline{e}.(P + Q) = \overline{e}.P + \overline{e}.Q\). Sequencing of waiting is also commutative and idempotent, so \(\overline{a.b} = \overline{b.a}\) and \(\overline{a.a} = \overline{a}\). Finally, a sequence of waits followed by the nullary process is itself nullary (i.e. any process where no firings are possible shares the same empty synchronization tree as any other process with that same property).

Some of these properties may be summarized by saying that idempotent, commutative monoid of sums of firings and waits in CEP acts as a left module over the idempotent, commutative monoid of sets of waits.

There are a few other, somewhat less common, properties of note. The first is absorbtion -- i.e. \(\overline{a}.P + P = P\), or more generally, when \(w_1\) and \(w_2\) are sets of waits with \(w_2 \subset w_1\), \(\overline{w_1}.P + \overline{w_2}.P = \overline{w_2}.P\). The second is a right decomposition. When \(ew\) is a sequence of firings and waits, and \(\overline{ew}\) is that same sequence, but with all firings transformed into waits, then \(ew.P = ew + \overline{ew}.P\). Right decomposition iterates to give a sort of ``simplicial decomposition'' such that \(a.b.c = a + \overline{a}.b + \overline{a.b}.c\) (and analogously for a sequence of firings of any size, or with interleaved intermediate waits). We may think of this as ``peeling off'' events one by one from the end of a sequence, and it is somewhat analogous to, e.g., single-static assignment semantics in imperative programming. 

\subsection{Normal Forms}

Simplicial decomposition lets us always reduce individual sequences of firings and waits in CEP to wait-normal form, as a (possibly empty) sequence of waits followed by a single firing. In turn, the algebraic properties of parallel composition let us reduce entire formulae into sums of sequences. Thus, this yields an ``elementary wait-normal rewrite'' that sends any term in CEP into an wait-normal flat form with extraneous waits absorbed.

One consequence of wait-normal form is that we have a syntactic criterion for a deadlock-free formula. Take a deadlock-free reduction to be one in which every remaining event is either excluded, or waiting on an event which is excluded, or a transitive extension thereof. Now, take a deadlock-free term in CEP to be one in which every possible reduction is deadlock-free. Then, it follows that a formula in CEP is deadlock-free if and only if it can be rewritten into a wait-normal form ordered such that no event occurs in a wait before it occurs in a firing.

The argument runs as follows: the prototypical example of a deadlock is \(\overline{a}.b + \overline{b}.a\). Obviously, in this case, it cannot be reordered so that no event in a wait occurs before a firing. Any addition of waits to the two components of the composition would have the same property. Therefore, to violate this property, we would need to introduce, e.g., a firing of \(a\) earlier in the formula.  However the only two ways for this firing to fail to occur (hence breaking the deadlock) would be if the prior occurrence  was itself deadlocked (hence breaking the property) or if it was excluded. But if the firing were excluded then \(b\) would be waiting transitively on an event which was excluded, and hence we would not be in deadlock.

This wait-normal form is, however, not yet normalized with regards to exclusions. To discuss this, we first need the notion of an expanded wait-normal rewrite. Consider the formula \(a + \overline{a}.b + \overline{b}.c\). This is clearly equivalent to \(a + \overline{a}.b + \overline{a.b}.c\), but the wait-normal rewrite has not yet resolved this. The expanded wait-normal rewrite is obtained by taking a formula in wait-normal form and then inductively ``expanding'' every wait. In particular, in each summand, for every wait \(w\) preceding a firing of each event \(e\), we create a new copy of the summand for each enablement of \(w\), which includes all the events that \(w\) itself waits on. When our wait-normal formula is deadlock free, this process terminates. As a slightly more complicated example, \(a_1 + a_2 + \overline{a_1}.b + \overline{a_2}.b + \overline{b}.c \) rewrites the final summand to \(\overline{a_1.b}.c + \overline{a_2.b}.c\).

Now, we can define a wait-and-exclusion-normal (henceforth wae-normal) formula as one that is in expanded wait-normal form, and furthermore such that no event firing (even with varying exclusions) is enabled by two sets of waits such that one is a subset of the other. 

Not every formula can be placed into wae-normal form. Take two chains of waits followed by the same event, \(\overline{w_1}.e![ex_1] + \overline{w_2}.e![ex_2]\), and \(w_2 \subseteq w_1\). In such a case, if \(ex_2 \subset ex_1\) then the additional exclusions in \(ex_2\) may be discarded, and absorbtion applies to normalize this to \(\overline{w_2}.e![ex_1]\). This is to say that if the second component is enabled (in terms of waits) whenever the first component is, and then it is always possible for the second component to fire without needing to take into account additional exclusions in the first. However, in the general case, when it is not the case that \(ex_2 \subset ex_1\),  this may give a situation where a context is invalid, but the addition of events makes it valid again. For example, \( \overline{a.b}.e + \overline{a}.e![c]\). It is not obvious what the correct way is to normalize this. In particular, while being wait-normal is a local condition, exclusion normality requires one to consider the whole system at once.

There is, however, a wae-normal rewrite that does not strictly preserve the synchronization tree. By example, it sends the problematic formula give above to \( \overline{a.b}.e_1 + \overline{a}.e_2![c] + \overline{e_1}.e + \overline{e_2}.e \). That is to say, when it encounters two chains of waits where one contains the other but the exclusions do not share this relationship, then it sends this to two distinct events, and furthermore it introduces two new ways to fire the original event. 

As there are now more events than before, the same synchronization trees cannot be produced. However, it is straightforward to verify that if we treat the new events introduced as ``silent'', then the synchronization tree of the original formula is the same as the synchronization tree of the second formula with the new events simply omitted. Thus, we nonetheless have a weak equivalence.

\subsection{The Correspondence to General Event Structures}

Translating between wae-normal formulae and general event structures is almost immediate. To each exclusive condition on a firing we generate a corresponding incompatibility in the consistency predicate that rules out the context given by the union of of the enabling conditions for that chain, the enabled event, and the excluded event. Then, to each fired event we associate the set of collections of its enabling events valid with regards to the consistency predicate, which yields the enabling relation. Because the wae-normality enforces the down-closed property of valid contexts, this suffices. We define the operation \(|P|\) as sending a wae-normal formula \(P\) in CEP to its corresponding general event structure. 

Translation from general event structures to wae-normal formulae in CEP is likewise mostly straightforward. For each minimal enabling context \(\{a,b,c\}\) of each event \(e\), we add a summand \(\overline{a.b.c}.e\). For each binary incompatibility, we add an appropriate exclusion to both incompatible elements. However, a complication arises when considering higher incompatibilities, such as the three-way incompatibility that allows \(a\) and \(c\) to occur, and \(b\) and \(c\) to occur, but not all three events. In such a case, these can be rendered into CEP through introduction of new ``virtual'' events to first choose which combination is possible and which is excluded. E.g., to express that \(c\) is exclusive of the joint firing of \(a\) and \(b\) we give the formula: \(ab + \overline{ab}.a + \overline{ab}.b + c![ab]\). Similar expansion can express all higher incompatibilities. We note that this sort of expansion means that the composition of translations to and from general event structures is far from the identity, and consider it future work to improve this situation.

\section{The Category of Package Repositories}
Since wae-normal formulae in CEP correspond to general event structures, building a category of general event structures (i.e. package repositories) should yield a notion of compositional categorical semantics, with operations in CEP corresponding to categorical operations. Further, other categorical operations may yield program transformations on CEP -- i.e., new admissible operators in the language. In turn, this all may shed light on which operations on package repositories, however represented, are particularly meaningful and useful.

To aid in the connection to intended concrete applications, in this section we will tend to refer to events as packages (with \(P\) in place of \(E\)), event structures as package repositories, etc. Context and dependency (i.e. enabling) information will often be collectively referred to as ``metadata''. We also will often denote package repositories by their corresponding normalized formula in CEP.

The usual way to give a category of event structures  considers partial functions between event sets, and imposes a partial injectivity condition \cite{winskel1986event}. 
This gives a setting in which events may be added or dropped, but neither merged nor split. But we wish to consider a wider set of transformations, in correspondence with the sorts of things people actually do to their packages. For example, frequently a package is refactored into a ``core'' and ``extended'' version, with the latter depending on the former, so that the core package can have faster build times and perhaps a smaller dependency footprint. For example, a library \texttt{web} might be refactored into \texttt{wev-core} containing the raw protocol interactions and \texttt{web-ui} containing utilities for generating html. Similarly, sometimes two packages providing related functionality are merged into the same package. This is to say that a third party may create \texttt{web-ui-forms}, extending the html library with better support for forms, and then at a later date this code may simply be merged back into the \texttt{web-ui} package.

Hence, we take the natural generalization to binary relations between event sets -- i.e. morphisms between sets \(s\) and \(t\) take the form \(R \subset s \times t = s \to \Pc(t)\). To every relation \(s R t\) there is an induced function \(\check{R} : \Pc(s) \to \Pc(t) = X \mapsto \cup_{x \in X} R(x)\), which acts on a set as taking the union of the elementwise action of the relation on its members. The move to binary relations is natural from another standpoint as well. We can consider a package repository as a sort of ``relationally fibered'' object, with valid enabling contexts ``living over'' the packages they enable. The natural notion of a morphism in an arrow category of relations is then itself relational. This has some similarity to the basic pairs used in formal topology \cite{sambin1998preview, bucalo2006completions}. 

\begin{definition}
A \textbf{relational morphism} between package repositories \((P, Con, \mapsto)\) and \((P', Con', \mapsto ')\) is a relation between package sets, given as a function \(f : P \to \Pc(P')\) (with the action on a set as the union of the elementwise action on its members) such that:
 \renewcommand{\labelenumi}{\roman{enumi}}
  \begin{enumerate}
   \item \(X \in Con \implies \check{f}(X) \in Con'\)
   \item \(X \vdash e \implies \forall e' \in f(e).\,  \check{f}(X) \vdash e'\) 
 \end{enumerate}

 (We state the second property with \(\vdash\) rather than \(\mapsto\) for notational convenience, but as discussed above they are equivalent formulations).
 \end{definition}

Unpacked, this means that relational morphisms between package repositories are given by arbitrary relations (i.e. multimaps) on their underlying package sets, subject to certain coherence conditions. The first condition ensures that any consistent context of packages relates to another consistent context -- i.e. if there is some consistent set of packages, and it is transported across the relation, it remains consistent. The second condition ensures that any enablement of a package by a context relates to an enablement of any related packages by the related context. So if a package is enabled in some context, then when that context is transported across the relation, the packages related are similarly remain enabled. 

This allows us to express the splittings and mergings as desired. As an example, the splitting of \texttt{web} is given as a morphism between a repository containing \(|\mathrm{web}|\) and  a repository containing \(|\mathrm{web\text{-}core} + \overline{\mathrm{web\text{-}core}}.\mathrm{web\text{-}ui}|\) with the mapping: \{web\} \(\mapsto\) \{web-core,web-ui\}.

Similarly, the merging of \texttt{web-ui} and \texttt{web-ui-forms} is given as a morphism between a repository which is given as \(|\mathrm{web\text{-}ui} + \overline{\mathrm{web\text{-}ui}}.\mathrm{web\text{-}ui\text{-}forms}|\) and a repository containing \(|\mathrm{web\text{-}ui}|\) with the mappings:  \{web-ui-forms\} \(\mapsto\) \{web-ui\}, \{web-ui\} \(\mapsto\) \{web-ui\}.

As two non-examples, we cannot take the identity morphism on packages and send \(|a|\) to \(|\overline{b}.a|\) nor to \(|a![b]|\). I.e. a package cannot become ``less buildable'' through a morphism.


Binary relations compose in the usual way -- i.e. \(g \circ f = x \mapsto \check{g}(f(x))\). It is also easy to see that the constraints we place on relational morphisms hold transitively, and that the identity relational morphism has the desired properties. Hence we have the following:

\begin{definition}
\(\PRb\) is the category with package objects given as repositories as morphisms given as relational morphisms.
\end{definition}

\subsection{Basic properties of \(\PRb\)}

A great deal of the structure of \(\PRb\) can be understood by considering its interaction with the category of sets with morphisms as relations, \(\Relb\). 

Every set (considered as a set of packages) gives rise to two basic repositories on it. The first is the free, codescrete, repository (denoted \(Codisc(S)\)), which has no incompatibilities and all packages enabled from any context. The second is the cofree, discrete, repository (denoted \(Disc(S)\)), in which only the empty context is valid, and no package is enabled by any context. As we shall see, both free and cofree repositories give subcategories of \(\PRb\) equivalent to \(\Relb\).

By abuse of notation, we also treat \(Codisc\) and \(Disc\) as endofunctors on \(\PRb\) which precompose to them the forgetful functor that sends repositories to their underlying package sets, throwing away their dependency and conflict information. For example, if we take \(A = a + b + \overline{a}.c![b]\) then \(Disc(A) = 0.a + 0.b![a] + 0.c![a,b]\) and \(Codisc(A) = a + b + c\). 

Free repositories, give \(\Relb\) as a (full and faithful) reflective subcategory of \(\PRb\). This is to say that any map of a repository into a codiscrete repository (considered as a free repository on a set) factors into the forgetful projection of a repository into its underlying set, followed by a relation of sets, as below:

\begin{equation*}
\begin{tikzcd}
A \arrow[rd, "f"] \arrow[d,"\musSharp{}"']      &           \\
Codisc(A) \arrow[r, "Codisc(f)"'] & Codisc(B)
\end{tikzcd}
\end{equation*}

 \(\Relb\) is also a (full and faithful) coreflective subcategory of \(\PRb\), since any map of a discrete repository into a repository factors into a relation on underlying sets of the two repositories followed by the addition of new dependency information and removal of conflicts, as below:
 
\begin{equation*}
\begin{tikzcd}
B                              &                                               \\
Disc(B) \arrow[u,"\musFlat{}"] & Disc(A) \arrow[lu, "f"'] \arrow[l, "Disc(f)"]
\end{tikzcd}
\end{equation*}

The reflector \musSharp{} and coreflector \musFlat{} give, respectively, the unit of an idempotent monad and counit of an idempotent comonad on \(\PRb\).


 
 
 
 
One key difference between \(\PRb\) and \(\Relb\) is that in the latter, all morphisms \(A \to B\) give rise to opposite morphisms \(B \to A\), as any relation \(R\) can be transposed into a relation in the opposite direction \(R^T\), and hence \(\Relb\) is self-dual. However, in \(\PRb\), this is not true. It is possible to remove incompatibilities, but not add them. It is possible to remove dependencies from dependency sets, but not add them, etc. Thus, for example, there are sixteen maps in \(\Relb\) between two two-element sets, but the only map possible from the codiscrete package repository on two elements (where every context is valid) to the discrete package repository on two elements (where only the empty context is valid) is the nullary map that sends every package to the empty set. In fact, one intuition for \(\PRb\) is that is can be developed by systematically breaking symmetries in \(\Relb\).

Developing on that, call a map which is identity on packages a ``metadata map''. Since metadata is part of the structure of package repositories as objects, a metadata map between two repositories, if it exists, is the unique such map. Furthermore, for any repository \(A\) there is a lattice of metadata maps which factorizes \(Disc(A) \to Codisc(A)\), and all metadata maps lie in such a lattice. It follows from this that any morphism \(A \to B\) factorizes uniquely as a relation on underlying packages that transforms metadata only (essentially) as induced by the underlying relation, followed by a metadata map which further removes conflicts, adds new possible dependency sets, etc. The ``essentially'' in the prior statement refers to a somewhat subtle point. When splitting a package into two, it may be the case that one of the two packages depends on the other. In such a case, that dependency must be induced in the same operation as the split, as introducing it later would be disallowed (since it makes a package ``less buildable'').

From this, one gets an intuition of \(\PRb\) as many copies of \(\Relb\) stacked on top of one another in a lattice. As one ascends from discrete repositories to ones with more possible build-plans, the possible morphisms at each level become restricted by the need to respect these plans. However, at a certain point, there is a ``phase transition'' and as almost every build-plan becomes possible, the addition of further plans \textit{allows} more morphisms, and so by the time we are in fully codiscrete repositories, we again have all arrows in \(\Relb\).

The properties of our (co)reflections give us intuition into certain properties of \(\PRb\). In particular, since a reflection creates all limits, limits in \(\Relb\) are created by the reflection from \(\PRb\). We can then confirm that the categorical product in \(\PRb\), just as in \(\Relb\), is disjoint union -- i.e. placing two repositories side by side without equating any packages. Coreflection likewise creates colimits and it is straightforward to check that categorical coproduct (\(\oplus\)), just as in \(\Relb\), is also disjoint union. As the ``nullary'' product and coproduct, the empty repository is thus both the initial and terminal object. Further, since \(\Relb\) does not have equalizers or coequalizers in general, then the same must hold for \(\PRb\).

For completeness, we record here the explicit construction of the coproduct:

\begin{definition}
In \(\PRb\) the coproduct \((P, Con, \mapsto) \oplus (P',Con', \mapsto ')\) is the package set \((P_{co},Con_{co},\mapsto_{co})\) where

  \begin{enumerate}
   \item \(P_{co} = P + P'\) (i.e a disjoint union of sets)
   \item \(Con_{co} = Con \times Con'\) (i.e. the combination of all valid contexts from either repository)
   \item \(p \in P_{co} \mapsto_{co} (c,c') \in Con_{co}\) iff \(p \in P \implies p \mapsto c\) and \(p \in P' \implies p \mapsto ' c'\)
   
  \end{enumerate}
\end{definition}

This yields the first element of a categorical semantics for CEP -- when \(P\) and \(Q\) do not share any events, then \(|P + Q| = |P| \oplus |Q|\). When events are shared, the situation becomes more complicated, and we will return to this question after further discussion of morphisms.

\subsection{Morphisms of Package Repositories}

We now consider classification of morphisms. Since faithful functors reflect monomorphisms, monomorphisms in \(\Relb\) come from monomorphisms in \(\PRb\), Hence monomorphisms in \(\PRb\) must act on the package sets of repositories just as monomorphisms in \(\Relb\). Further it turns out that it suffices to check only this condition, since the additional conditions on \(\PRb\) rule out some morphisms, but do not otherwise quotient or relate them. In particular, a morphism \(f : a \to \Pc(b)\) is a mono in \(\Relb\) (and hence \(\PRb\)) if and only if \(\check{f} : \Pc(a) \to \Pc(b) = x \in \Pc(a) \mapsto {\bigcup}_{e \in x} f(e)\) is a monomorphism in \(\Setb\), i.e. an injective function. Epimorphisms arise dually, and are perhaps easiest thought of as the relational inverses (transposes) of monomorphisms.


Monomorphisms capture a very broad notion of extension of package repositories. Consider the free repository with two packages \(\{a,b\}\), and the monomorphic map into a repository with three packages \(\{d,e,f\}\), such that \(a \mapsto \{d,e\}, b \mapsto \{e,f\}\). This may reflect a situation where two packages have been refactored to extract out a common subpackage. For example, take \(a\) and \(b\) to be two libraries for list manipulation that both define the \texttt{split} function, and the monomorphism to be the refactoring that extracts a \texttt{split} library on which they both depend.

Similarly epimorphisms in \(\PRb\) are expressive enough to describe situations when two packages have been merged into a single package, or when a common library has had various components ``inlined'' into multiple packages (i.e. a converse refactor to that above, in which the split library is removed, and each of the list libraries again inlines the function). The prior examples of splitting and merging \texttt{web} libraries are thus monic, and epic, respectively.
   
Notably, any morphism which is identity on packages is both monic and epic. However, since such morphisms may remove conflicts or add new possible dependency sets, they are not necessarily isomorphisms, and as such, then \(\PRb\) is not a balanced category. Rather, monic epics in \(\PRb\) are precisely the ``metadata maps'' discussed earlier. This analysis relates to real-world tools, as the Hackage repository of Haskell packages allows a ``metadata-revision'' operation which does not touch the existing package set, but does alter dependency information. Here we see that ``safe'' (in the sense of not breaking existing builds) revisions to metadata  are precisely those which are valid monic epics in \(\PRb\).

As a whole, these various sorts of morphisms give a rich structure on ``safe'' (in the sense of not breaking existing builds) transformations of package repositories, most of which are not implemented in existing tools. Of course, we are interested in modeling also the single operation that \textit{does} generally exist -- addition of a single new package. Here we turn to a stronger notion of injection, namely a split monomorphism, i.e. a map \(s\) such that there exists a retract \(r\) with \(r \circ s = id\). In \(\PRb\), as in \(\Relb\), a split monomorphism acts on packages by sending them to unique, nonoverlapping, package sets. Further, it is clear that a split monomorphism cannot remove conflicts between existing packages, and cannot add new possible dependency sets to existing packages, as both operations are ``one way''. 

To understand the difference between a split monomorphism and a general monomorphism, consider the example above where two list libraries (\(a\),\(b\)) have had a splitting library, \(e\) factored out. If I depended on library \(a\) before, then I would need to depend on \(a\) and \(e\) afterwards. But if the converse epimorphism that merged \(e\) back into \(a\) and \(b\) was performed, there is not enough information to know if only some, or all of the functionality of \(e\) was sent to \(a\) and \(b\) respectively. Accordingly, I would then need to depend on both \(a\) and \(b\) after the merge. Hence the common sublibrary refactor, while a monomorphism, is not a split monomorphism. On the other hand, another running example--the extraction of \texttt{web-core}--is invertible such that the round trip is identity, because the splitting of the \texttt{web} package does not overlap with splitting anything else. In that case then, there is a split monomorphism.

We call a split monomorphism minimal if the only way to further factor it into a composition of split monomorphisms is if one of the two monomorphisms it factors into is an isomorphism. Thus, a minimal split monomorphism in \(\PRb\) is either the extension of a package repository by a single new package (i.e. the basic operation on existing package repositories) or the splitting of a package into two packages, both sharing the same dependency sets and conflicts as the original. 

An extension can then be distinguished from a splitting by characterizing it as a minimal split monomorphism which is also a function on sets (i.e. relates each element in its domain to precisely one in its range). Extensions compose to further extensions, and so we recover the usual notion of an injective function.


The same conditions give us the retract of a split monomorphism as a split epimorphism, which corresponds to the deletion of a package that nothing depends on, or the merging of two packages with the same dependency and conflict information. The preceding suggests that existing tooling for package repositories may want to implement some of the three operations not usually present (``safe'' deletion, splitting, and merging) as they all seem like basic mathematical primitives that are easy to understand, and perhaps usefully expressive.

In any case, this provides our second element of a categorical semantics for CEP -- \(|P + \overline{e_1.e_2..}.d![a_1,a2..] + \overline{e_3.e_4..}.d ..|\), where \(d\) does not appear in \(P\) corresponds to a split monomorphism from \(|P|\) to \(|P|\) extended by a single new package as well as relevant dependency sets and conflicts.


\subsection{Pushouts and gluing of repositories}

We have given semantics to the situation of concurrent composition where no variables are shared, and to the situation where dependency and conflict variables are shared in one direction but not event variables. We now turn to the general construction which should underlie giving semantics to composition in general. 

The fundamental tool in composing repositories will be the categorical pushout operation. The coproduct operation of placing two repositories side-by-side is already an instance of this, as it can be constructed as the pushout over the initial object. Recall that a pushout in a category is computed as the colimit of a span \(A \leftarrow C \rightarrow B\), consisting of an object \(X\) and maps \(A \rightarrow X \leftarrow B\), such that any other \(Y\) of the same shape factors through \(X\), as pictured below:

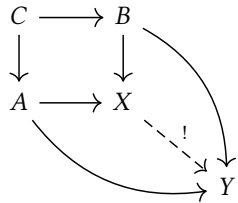
\begin{equation*}
\begin{tikzcd}
C \arrow[d] \arrow[r]               & B \arrow[d] \arrow[rdd, bend left] &   \\
A \arrow[r] \arrow[rrd, bend right] & X \arrow[rd, "!", dashed]          &   \\
                                    &                                    & Y
\end{tikzcd}
\end{equation*}

We do not expect pushouts to exist in general in \(\PRb\). However, when they do exist, this means there is a unique best way to merge two repositories with regards to the data they share. For example, in \(\Setb\), where pushouts always exist, when the pushout of a span of injections \(A \hookleftarrow C \hookrightarrow B\) consists of the disjoint union of \(A\) and \(B\) with elements identified when they both come from the same element of \(C\). E.g., the pushout of the span (with \(A\), \(B\) and \(C\) disjoint) \(A \cup C \hookleftarrow C \hookrightarrow C \cup B\) is the set \(A \cup C \cup B\). Thus we can speak of such a pushout as a ``gluing along C''. 

We consider now the analogous situation in \(\PRb\) -- a pushout of a span of relations that are injective functions. As \(\Relb\) is left adjoint to \(\Setb\) it preserves all colimits. Thus in \(Rel\) the pushout of injective functions \(f, g : A \hookleftarrow C \hookrightarrow B\) works the same way as in \(\Setb\) -- i.e. it is the disjoint union of \(A\) and \(B\) with elements identified when they are both related to the same element in \(C\). Since injective functions in \(\PRb\) cannot affect existing dependency or conflict data, this suffices to characterize pushouts of injective functions in \(\PRb\) as well.

Furthermore, we have that the pushout of two epi-monos (metadata morphisms that are identity on packages) in \(\PRb\) also always exists, in the obvious way -- i.e. new build plans introduced in either leg are present in the pushout, and newly valid contexts in either leg are valid in the pushout. 


By the pushout pasting law, we then have that a span of injections of the discrete repository into two repositories has a pushout, since we may factor such injections into an injection between discrete repositories (which is a strict mono) followed by an epi-mono.

Thus, given two repositories which we wish to compose, we can take span given by the discrete repository of the events which are to be identified in each, with injections to each repository, and then take the pushout of this.

For completeness we describe here the explicit construction of a pushout along injections of shared packages.

\begin{definition}
In \(\PRb\) the pushout \((P, Con, \mapsto) \boxplus (P',Con', \mapsto ')\) along the injections of the set of their shared names, \(N\) is the package repository \((P_{\boxplus},Con_{\boxplus},\mapsto_{\boxplus})\) where

  \begin{enumerate}
   \item \(\sim{N}\) is the relation that couples elements of \(P\) and \(P'\) when they are identified by \(N\).
   \item \(P_{\boxplus} = P+P'/\sim{N}\) (i.e a pushout of sets)
   \item \(Con_{\boxplus} = Con \times Con' / \sim{N}\) (i.e. any valid context remains valid)
   \item \(p \in P_{\boxplus} \mapsto_{\boxplus} con \in Con_{\boxplus}\) iff \(p \in P \implies p \mapsto pr_1(con)\) or \(p \in P' \implies p \mapsto pr_2(con)\)
  \end{enumerate}
\end{definition}

This yields the general categorical semantics for parallel composition in CEP. When \(P + Q\) is wae-normal, \(|P + Q| = |P| \boxplus |Q|\) -- the pushout of the two processes along their shared names.


The machinery we have built is more general than such a case, and this extra generality should be of some interest. In particular, a common situation in package management is the existence of a branching tree of package repositories. Along with a socially-agreed-on ``central'' repository for the language, individual projects may host their own repositories, and commercial enterprises may host ``in-house'' repositories not accessible to outsiders, etc. The typical approach of package managers is to just treat these as a linearized list, with information from each successive repository superseding information from repositories earlier in the list. But such a treatment does not necessarily produce a valid repository. However, because we know that both repositories have evolved from a common ``base'', then there exist injections from this base to both repositories, and the pushout along these can yield a proper composition.


\section{Applications}

We sketch here some applications of the framework we have developed in modeling more particular sorts of situations and questions that arise in the treatment of package repositories and the evolution of package versions.

\subsection{Package Version Policies}

The first modern notion of a package version policy was introduced in 2005 by Sven Moritz Hallberg, in the article ``Eternal Compatibility in Theory'' \cite{ect} (where it was considered an inferior alternative to a more complicated scheme). It was later proposed more directly for Haskell in 2006 \cite{librarypolicy}, and over time gained social acceptance. A few years later, the beta semantic versioning spec \cite{semver} was released, and began to gain traction across a broader range of languages. The motivation behind various package version policies is broadly similar -- as libraries began to be not bundled with software (``vendored''), but to depend on one another with package repositories supplying them to end-users as demanded, it became imperative that they not declare dependencies on specific versions of other libraries, but rather a range of them. This in turn deferred to the end-user (or their package management software, rather) the responsibility to assemble a coherent choice of versions such that various libraries could work together. 

The solution provided is also broadly similar -- that authors of packages distinguish between ``major'' (so-called ``API-breaking'') changes and ``minor'' changes (i.e. ones conforming to some definition of ``backwards-compatible''). When a package is modified in a way such that the API surface changes incompatibly (so that something that built against it in the past may fail to do so), then the major version is incremented. For example, a function deletion results in a major change, since everything that used the function now fails. When a package is modified in a way such that the API changes ``compatibly'' then the minor version is incremented. For example, under the assumption that users import functions from modules explicitly, a function may be added in a minor version, since this would cause no code that built before to now fail. Version policies are now pervasive enough that special syntax for encouraging their use exists in many formats of package metadata. For example, as discussed earlier, in \texttt{npm} metadata, ``\^{}1.2'' refers to the range of minor versions greater than or equal to 1.1 and less than 2. In \texttt{cabal} file syntax, ``\^{}>= 1.1.0'' plays a similar role.

Since there are no rules relating major versions of a package, there is no particular semantic treatment to give them, and it makes sense to consider major versions of a package part of its name, rather than its versioning information. However, the categorical semantics we have produced can shed light on minor versions. What exactly constitutes a breaking change or not may vary between languages, and conventions within languages, and thus is a matter of some contention. For the purposes of this paper, we consider only the situation where an increment to a minor version genuinely means that anything which built against the earlier version can build against the later version. Notationally, we take \(c[p := q]\) to represent the context \(c\) with any occurrence of \(p\) substituted for \(q\).

\begin{definition}
In a package repository \((P, Con, \vdash)\)  two packages \(p\) and \(q\) are version-related when \(\forall c \in Con. \implies c[p := q] \in Con\) and furthermore, \(\forall a \in p. c \vdash a \implies c[p := q] \vdash a\). In such a case, we may say that \(p\) is a lower version of \(q\).
\end{definition}

From the above definition, it is immediate that every versioning relation on a package repository \(P\) such that \(p\) is a lower version of \(q\) induces a valid (idempotent) endomorphism sending the lower version of an element to the higher version. This factors into two morphisms -- the mapping into a repository with \(p\) merged into \(q\),which is a minimal epimorphism (not necessarily split) that is also a function on sets, followed by the (split) injection back into the original repository. Thus, in a sense, the contraction of versions is a sort of symmetric dual to the addition of a package. Since pushouts exist as well for mergings of packages (since they act as a function on sets), then it follows that version-relations compose nicely with regards to categorical structure, and with regards to the algebraic structure of CEP.

Finally, we note that it follows from the foregoing that there exists a build plan for a package in \(P\) iff there is a build plan for that package in \(P[p:=q]\). We believe this helps to explain and account for the importance of version policies in large-scale package repositories -- they provide a way to systematically reduce the complexity of a system of dependencies while preserving its essential properties.


\subsection{Granular Updates of Single Libraries}
The same analytical tools that help us to understand dependencies between entire packages in package repositories can be applied more granularly to understand the slow process of backwards-compatible API changes to a single library. One motivating example here is the ``three release'' policy of GHC, which mandates that any change to the \texttt{Prelude} module imported by all packages must be done in such a way that code can be written to work with the latest three releases of the GHC compiler.


Here, rather than consider a package as a single unit, we can instead consider it as a portion of a ``repository'' itself, that is to say a collection of individual functions, some only ``enabled'' to be defined when others have already been brought into scope. This is to say we take a monomorphism from a package repository into another repository with a chosen package (considered here as a single module for simplicity's sake) split into the dependency structure of its individual functions. In such a case we can then consider those morphisms which only act in an interesting way on the ``exploded'' package. From this, it follows that addition of a new function is always allowed, but deletion of a function is only safe if if nothing depends on it.

We can then, given some desired end-state, factor the morphism we desire into a series of alternating morphisms that affect only the exploded package or only the wider ecosystem. As a simple example, the renaming of a function can be accomplished first by addition of a new function with the new name, then (typically signaled by a deprecation warning), a morphism which sends packages depending on the old function to packages depending on the new function, and finally a morphism which drops the old function, as nothing now depends on it. More complicated examples can be treated in a similar fashion.

This idea of ``exploding'' a single package into its tree of individual functions and their dependencies has been proposed in practice as a way to handle versioning and upgrades by Joe Armstrong \cite{armstrong}
and explored by languages such as Unison and Dhall \cite{unison, dhall}.

The formalism developed here lets us articulate the relationship between these sorts of approaches and more traditional module- and package- centric mechanisms.

\subsection{Language-Specific Semantics of Conflict}
As discussed earlier, different package managers for various language ecosystems have different semantics as to when conflict relations arise between packages, typically due to different versions of a package being required by different dependencies -- otherwise known as the diamond dependency problem. The approach given here allows us to compare these by modeling them as alternate reduction semantics for CEP. Here we sketch how that would work.

 The Haskell model is the most straightforward -- there is a rule in the cabal solver that no two versions of a package may be in scope at once, full stop. The solver tries to find a path such that all desired packages can be installed at once in a conflict-free way, or fails. This is the standard reduction semantics.

In \texttt{npm2} semantics, every dependency is vendored, which is to say that it is brought into scope solely for the use of the package which depends on it. \cite{npmworks} Hence, in our example with \texttt{leftpad} and \texttt{text}, if we assume \texttt{text-1.1} was installed first, and then \texttt{leftpad-1.2} depending on \texttt{text-1.2} was installed, we would have a trace of \texttt{[text-1.1,text-1.2\textsubscript{text-1.1},leftpad-1.2]}. 

Java, meanwhile ``solves'' the problem by letting only the most recently loaded version of a package ``win''. So the same situation would lead to a trace \texttt{[text-1.2,leftpad-1.2]}, with \texttt{text-1.1} nowhere to be found. 

Finally, Rust solves the problem by ``mangling'' names where necessary to disambiguate claims to the same namespace. So on the same example again, we would obtain a trace \texttt{[text-1.1, mangle(text-1.2), leftpad-1.2\textsubscript{\textit{mangle}(text-1.2)}]}. \cite{rusthell}.

\subsection{Package Dependency Solving}

The CEP formalism also immediately gives an algebraic characterization of what it means to ``solve the dependencies'' of a package when a program such as \texttt{cabal} is asked to install it. In particular, given a deadlock-free CEP formula representing a package repository, a package has a build plan (i.e. the dependencies can be solved) precisely when that formula, when placed in wae-normal form, has a wait-set which enables that package such that no two elements within the set are in exclusive conflict. This is of course far from an efficient algorithm (as wae-normal form can be much larger than the input data), and indeed no particularly efficient algorithm can be found, as the problem (as discussed in the introduction) is NP-complete. 

However, framing things in these terms is of some use, because it opens the way to domain-specific efficiencies that could be derived from various formal constructions and algebraic properties of the underlying system. As a simple example, given some package-version information, we can cut down the search space necessary to consider. When two packages are version related, any dependency problem solvable by considering the full repository is also solvable when considering the repository with all terms concerning the lower package omitted. Thus we can always quotient a package repository by all version-relations. 

Further, we observe that if a package has solvable dependencies in either \(P\) or \(Q\), then it has solvable dependencies in \(P + Q\). This suggests an approach similar to that used in \cite{master2020open}. In particular, one should be able to compute for the pushout itself what new build-plans it enables, and then ``glue'' the build-plans along  this path. Solving the dependencies of an event can then also be seen as constructing a minimal subformula containing a firing of that event, in which every transitive wait is also enabled, and in which no two events are in conflict. In this formulation, a dynamic programming approach is immediately evident.

All of the foregoing is of course very well known, just restated in somewhat different language. In particular, it gives a correspondence, in the case of no conflicts, to the problem of solving a goal in classical horn logic (as each pairing of a wait-set and event corresponds to a clause in implicative normal form). The introduction of conflicts then yields a particular sort of negation, which seems distinct from the standard approach of negation as failure. One approach would be the extension of mixed forwards- and backwards-chaining techniques to this setting \cite{harland2000forward}. 

\section{Related and Future Work}

As discussed in the introduction, apart from \cite{di2006edos, abate2012dependency}, there is very little in the way of formal study of package management.  Dependency structures also play a heavy, although largely implicit role in the formal treatment of build systems as studied in \cite{mitchell:shake_24_sep_2018}. Our purpose in this paper is not so much to ``settle'' any major questions, but simply connect the question up to well-known approaches to other aspects of programming languages and so lay the basis for future work. 

Similar constructions to those on package repositories also occur in studying the evolution of repositories in version control systems. A programming language approach to the semantics of version control has been studied in \cite{swierstra2014semantics}. Version control and attendant topological models have been studied in \cite{angiuli2014homotopical} and \cite{mimram2013categorical}. Further, we suspect that the obstructions to general (co)products in \(\Relb\) and \(\PRb\) are similar to the obstructions that prevent patches from generally commuting.

Most directly, we have drawn on the theory of general event structures as developed in \cite{winskel1986event}. As that paper describes, general event structures do not immediately take semantics in domains. Rather, it gives a mapping from so-called ``prime'' event structures, which are general event structures such that every event has a single minimal enabling context (i.e. a ``prime'' cause). Hence, general-event structures as such have not been much-studied. Recent work in \cite{baldan2017domains} has picked out a subclass of connected event structures, more general than prime event structures, and which has a ``splitting'' coreflection similar in spirit to that we give for the wae-normal rewrite.

Our CEP process calculus takes inspiration from the process calculus developed there for prime event structures. The ensuing literature on concurrency, event structures, and various forms of bisimulation and trace semantics is far too rich to survey here. Typically, process calculi such as the \(\pi\)-calculus \cite{milner1992calculus} do not have an explicit notation for conflict, and instead allow it to emerge from nondeterministic choice. Here, we instead recover nondeterministic choice from explicit conflict. 

On the categorical side, we have given a new category for event structures that considers not merely partial functions, but relations. This extra generality lets us model a wider range of refactoring operations, which can be considered as between packages, or ``within'' a package viewed as an exploded dependency structure. Further, the lack of general coproducts highlights important issues in what sorts of operations can and cannot compose. 

The long-term goal in this line of work is an understanding of dependency and conflict from a topological standpoint. One approach is through the order-theoretic properties of the configuration family induced by event structures \cite{bazerman2020topological}. The work here suggests at least three other possibilities. First, the connection of the category \(\PRb\) to the structure of basic pairs used in formal topology \cite{sambin1998preview}. Second, the possibility of extending the discrete and codiscrete structure with some further form of ``cohesion'' \cite{lawvere2007axiomatic}. Third, pursuing the algebraic connection of CEP formulae to a module over a monoid as a setting for directly conducting homological algebra \cite{anderson2012rings}. The connection between concurrency theory and topological structure has also been explored through directed algebraic topology as in as in \cite{fajstrup2016directed}.

On the practical side, we would like to set the formalisms here to work in future work on package management systems. As discussed, pushout constructions could be of use in specifying the correct way to compose multiple repositories into a single object for use by a package management tool. Further, it may be worthwhile to directly endow repositories with constructions representing splitting of merging of packages. And not least, we would like to develop a correct-by-construction package dependency solver, and explore domain-specific efficiencies that could be derived from various formal constructions.

Also, we hope that the general approach of modeling dependency relationships through event structures may be of use in mechanized representations of knowledge (especially mathematical knowledge), such as that pursed by the Formal Abstracts project \cite{fabstract}. 

\section{Conclusion}
We have described the problem of package management in terms familiar to programming language researchers, showing how a package repository may be interpreted as the declarative specification of a concurrent program. Further, we have specified a process calculus corresponding to this model, and explored its reduction semantics and some of its algebraic properties. Making use of these properties, we then constructed a general category of package repositories in which normal-form expressions in the process calculus could take semantics. In more purely mathematical terms, this also means we have developed the notation of an algebraic presentation of a general event structure. Finally, we have surveyed some potential applications of this formalism in studying version policies and library updates, as well as the conflict resolution semantics of various package management systems.

\begin{acks}
The initial impetus for this research grew out of many interesting conversations with Herbert Valerio Riedel regarding package repository semantics. The need to render these ideas more formal was due to the careful interrogation of Simon Peyton-Jones. Jonas Frey first pointed out to us the relationship to event structures. Many ideas related to this work have been developed in close collaboration with Raymond Puzio. Many helpful comments on this paper were contributed by Jeff Polakow. We also thank Brendan Fong for setting us straight on some categorical idiocies.
\end{acks}

 \bibliographystyle{amsalpha}
 \bibliography{repos-ppdp}

\end{document}